# Graphene coating of Nafion® membranes for enhanced fuel cell performance

Jasper Ruhkopf[1,2], Ulrich Plachetka[1], Michael Moeller[1], Oliver Pasdag[3], Ivan Radev[3], Volker Peinecke[3], Marco Hepp[4], Christian Wiktor[4], Martin R. Lohe[5], Xinliang Feng[6], Benjamin Butz[4], Max C. Lemme[1,2,*]

[1]AMO GmbH, Otto-Blumenthal-Str. 25. 52074 Aachen, Germany
[2]RWTH Aachen University, Chair of Electronic Devices, Otto-Blumenthal-Str. 2, 52074 Aachen, Germany
[3]The Hydrogen and Fuel Cell Center - ZBT GmbH, Carl-Benz-Str. 201, 47057 Duisburg, Germany
[4]University of Siegen, Micro- and Nanoanalytics Group, Paul-Bonatz-Straße 9-11, 57076 Siegen, Germany
[5]Sixonia Tech GmbH, Maria-Reiche-Str. 3, 01109 Dresden, Germany
[6]Technische Universität Dresden, Chair for Molecular Functional Materials, Mommsenstr. 4, 01062 Dresden, Germany

*Corresponding author: lemme@amo.de

## Abstract

Electrochemically exfoliated graphene (e-G) thin films on Nafion® membranes exhibit a selective barrier effect against undesirable fuel crossover. The approach combines the high proton conductivity of state-of-the-art Nafion® and the ability of e-G layers to effectively block the transport of methanol and hydrogen. Nafion® membranes are coated with aqueous dispersions of e-G on the anode side, making use of a facile and scalable spray process. Scanning transmission electron microscopy (STEM) and electron energy-loss spectroscopy (EELS) confirm the formation of a dense percolated graphene flake network which acts as diffusion barrier. The maximum power density in direct methanol fuel cell (DMFC) operation with e-G coated Nafion® N115 is 3.9 times higher than the Nafion® N115 reference (39 vs. 10 mW $cm^{-2}$ @ 0.3 V) at 5M methanol feed concentration. This suggests the application of e-G coated Nafion® membranes for portable DMFCs, where the use of highly concentrated methanol is desirable.

Proton exchange membrane fuel cells (PEMFC) are an emerging technology to realize equal-zero-emission for transportation powered by gaseous hydrogen [1]. Due to their high power density and light weight, they are ideally suited for automotive applications. In addition to PEMFCs, direct methanol fuel cells (DMFC) are available, that are driven by water-diluted methanol [2]. DMFCs are considered as a potential next-generation power supply technology for off-grid applications or portable electronics, like cellphones or laptops. Complexity, high production costs and open technological challenges prevent a wide distribution of fuel cell technology so far. Several shortcomings of the state-of-the-art technology originate from the membrane material used, and the ideal material has yet to be invented. The perfect PEMFC and DMFC membrane would completely block any other ions and molecules from passing from the anode to the cathode side and vice versa, while providing maximum proton conductivity. Perfluorosulfonic acid (PFSA)-based membrane materials, e.g. Nafion®, Aquivion®, 3M™, etc. are established as state-of-the-art, because they possess high proton conductivity and offer excellent mechanical and chemical stability. However, their proton selectivity is far from being ideal and limits the overall fuel cell performance. A significant amount of fuel molecules (e.g. hydrogen or methanol) can cross these membranes and lead to parasitic chemical redox reactions on the cathode side, which lowers the achievable cell voltage, power density and cell lifetime due to cathode degradation [3]. Particularly PEMFC and DMFCs suffer from this problem, also known as hydrogen or methanol crossover, respectively. Research focusses on several different approaches to reduce the crossover, while simultaneously retaining a high proton conductivity. The most obvious method is to use a thicker membrane of the same proven PFSA material type, which in fact lowers the crossover, but also reduces the proton conductivity, respectively power density significantly [4]. Research has also been directed at finding a polymeric replacement material. Polymers like polystyrene-sulfonic acid (PSSA), sulfonated polyimide (SPI) or sulfonated polyether ether ketone (s-PEEK) have been proposed, but do not reach the same overall performance and durability as PFSA-based membranes [5]. Recently, composite materials, consisting of a PFSA matrix and filler materials are explored extensively. Different types of clay, graphene oxide (GO) and other two-dimensional materials like hexagonal Boron Nitride (h-BN) or MXenes (2D inorganic compounds) have been studied as filler material in PFSA-composite membranes [6–9]. While the latter approach is effective in reducing the crossover, the increased complexity of the associated membrane fabrication process with low manufacturing readiness levels prevents the approach from being used in commercial applications. GO membranes fabricated by vacuum-filtration have also shown reduction of fuel crossover compared to Nafion®. However proton conductivity was reduced as well [10,11]. In addition, the GO films are not mechanically stable as free-standing membranes and require further support by Nafion® membranes to withstand operation in a fuel cell setup. To overcome those issues, crosslinked sulfonated GO coatings applied on a Nafion® membrane were explored [12], however, data for high methanol concentrations are not available. Attempts have been made to use thin spray-coated GO layers, directly applied on the fuel cell electrode as PEM replacement [13]. While the layers showed competitive through-plane proton conductivity, the measured crossover was not notably reduced compared to a Nafion® reference. Comprehensive reviews of the use of graphene-based materials for fuel cell applications and of polymer nanohybrid membranes can be found in references [14] and [15], respectively.

In this work, we present a facile method to reduce the fuel crossover in PFSA membranes by applying a functional graphene-based film as a selective diffusion barrier on the Nafion® surface by spray coating. The approach combines the advantages of Nafion® and graphene in a comparably simple, inexpensive and scalable process. Graphene dispersions with various flake sizes are evaluated regarding their suitability as a proton selective coating material and optimized to minimize losses in proton conductivity. Structural analyses of the coatings and membrane cross-sections are carried out by scanning electron microscopy (SEM), scanning transmission electron microscopy (STEM) in combination with electron energy-loss spectroscopy (EELS) and atomic force microscopy (AFM). Further characterization is performed by Raman spectroscopy. The permeability of the coated membranes is characterized with a mass spectrometer setup (MS) for hydrogen / PEMFCs and with a liquid methanol sensor (LMS) for methanol / DMFCs. Finally, fuel cell operation is tested in DMFC configuration with membrane electrode assemblies (MEA).

## Results and discussion

Nafion® membranes were coated with dense, continuous graphene thin films by spray-coating of electrochemically exfoliated graphene-based dispersions (see **Figure 1a** and experimental methods section for details). In this study, the average graphene flake sizes and applied graphene loadings were varied as experimental parameters. After characterizing the coated membranes, MEAs were prepared for methanol permeability experiments with the graphene diffusion barrier applied to the anode-side of the membrane, which was sandwiched between Nafion® membrane and anode catalyst (**Figure 1b**). The corresponding measurement setups for determining proton conductivity, permeability to hydrogen and methanol, proton conductivity and performance in DMFC operation are depicted in **Figure S1**

Top-view SEM and AFM of the as-deposited graphene layers on the Nafion® membranes confirmed the high coating quality after optimization yielding continuous and flat graphene films. **Figure 2a** exemplarily shows such an optimized graphene film fabricated from an aqueous dispersion containing 1 µm sized graphene flakes. The graphene loading was estimated to be 130 µg $cm^{-2}$. The flakes formed a randomly stacked, percolated network when spray-deposited on the substrate, in line with previous reports [16]. The resulting layer was quite smooth, which impedes the recognition of individual flakes through top-view SEM imaging. This is the result of carefully optimized processing conditions and choice of materials, while earlier attempts resulted in far less ordered graphene coatings (compare **Figure S2**).

Atomic force microscopy was performed on a Nafion® HP membrane before and after coating. The Nafion® HP reference sample had a RMS roughness of 5 nm over a scanning area of 10 µm x 10 µm. After coating, the roughness slightly increased to an RMS value of 19 nm, which corroborates the flat alignment of the flakes on the Nafion®, indicated by SEM (**Figure 2a**, inset).

Cross-sectional characterization by optical microscopy (OM) and TEM was conducted on the graphene coated Nafion® HP substrates for a more detailed analysis of the geometrical flake arrangement. The corresponding optical micrograph in **Figure 2b** confirms a well adhering, homogeneously thick and continuous graphene layer (goldish color) on top of the Nafion® without gaps within the resolution limitation of the applied optical technique. In annular dark-field mode STEM (ADF-STEM) (inset of **Figure 2b**) the graphene film appears as a bright and homogeneous contrast with few minor flaws like small pores or remaining inclusions from the graphene ink, as indicated by black arrows. The imaged region featured an unusually high density of pores or inclusions and was chosen to illustrate the type of flaws in the otherwise smooth and dense film. The actual nature and impact of individual flaws in the graphene film is difficult to identify and beyond the scope of this publication. A detailed view of the graphene flake network at higher magnification is shown in **Figure 2c**. Single flakes, aligned parallel to the electron beam, can be distinguished as bright lines.

This interpretation of the contrasts in STEM imaging is confirmed by qualitative STEM-EELS analyses. **Figure 2d** shows the graphene distribution with a high degree of $sp^2$-hybridized carbon in comparison to the respective ADF-STEM image. The corresponding Nafion® and epoxy maps are shown in **Figure S3**. In contrast, similar data from the early coating attempt clearly shows severe structural disorder and resulting porosity (**Figure S2)**. The reference spectra of each component, extracted from the mapping, are presented in **Figure 2e**. The graphene reference (red) shows a distinct π* peak at 285 eV, typical for $sp^2$-hybridized carbon. While this peak is also present in the other two reference spectra it is less pronounced and broader, indicating the expected predominant $sp^3$-hybridization of carbon in the epoxy resin and Nafion®. The fit to the epoxy resin reference not only features increased intensity in the according layer of the TEM sample, but also in between Nafion® and the graphene film. This is attributed to slight local delamination of the graphene film and intrusion of the liquid epoxy resin during TEM sample preparation. This delamination is rather unlikely during operation of the assembled cell though, since the individual layers are compressed by the bipolar plates of the complete MEA assembly. Additional Raman data confirms the graphitic nature of the thin films (**Figure S4)**.

The fuel permeability of the e-G films and their performance as barrier layer in DMFC operation was studied. The hydrogen permeation through coated and uncoated reference Nafion® HP membranes under different hydration conditions was measured with an *ex-situ* mass spectrometer setup. In dry condition at 80°C, the graphene coating reduced the permeation by a factor of 39 from 1274 ppm (reference) to 32 ppm. The permeation reduction factor of 29 was maintained at 80°C and a relative humidity of 65 % (2151 ppm compared with 73 ppm). Further experiments focus on DMFC performance and methanol permeability.

Coated Nafion® N115 with various graphene loadings and bare Nafion® reference samples were assembled into MEAs and characterized in a DMFC test setup. I-V-curves were recorded for 1M and 5M concentration of methanol (**Figure 3a**). The sample with the least graphene loading (32.5 µg $cm^{-2}$) showed a similar performance at 1M methanol as a bare N115 reference sample (**Figure 3a**, green and black data, respectively). Higher graphene loadings lead to a poorer performance at 1M methanol compared to the reference, indicated by the early breakdown of the output potential [V] at lower current densities. This is attributed to a progressive reduction of proton conductivity by the additional graphene layers. At 5M, the MEA with 32.5 µg $cm^{-2}$ loading performed worse than the reference. Increasing the graphene loading, in contrast, significantly increased the breakdown current density for 5M methanol concentration. Both 65 and 130 µg $cm^{-2}$ graphene-Nafion® N115 samples showed better DMFC performance in terms of breakdown current compared to the non-coated Nafion® reference. In fact, the 130 µg $cm^{-2}$ sample showed a comparable performance at 5M as the reference at 1M. This performance improvement through graphene coatings with higher loading can be explained by a decrease of methanol permeability while preserving high proton conductivity of the MEA.

The open circuit potential (OCP) or open circuit voltage (OCV) of a cell and its deviation from the theoretical reversible cell potential of 1.21 V are a figure of merit to quantify methanol crossover [17]. OCP values for various graphene loadings were extracted from the I-V-curves. The graphene samples with 65 and 130 µg $cm^{-2}$ loading exhibit an increased OCP of up to 0.67 V at 5M methanol concentration compared to a value below 0.45 V for the reference, which indicates a reduction of methanol crossover by the graphene coating. The methanol barrier effect was further studied in permeability experiments on the MEAs, which were performed by a novel measuring technique (details can be found in the experimental methods section). The measurement was carried out in the DMFC test setup under *ex-situ* operating conditions in equilibrium state. The concentration of the permeating methanol in the cathode outlet was monitored by a liquid methanol sensor. The respective methanol permeabilities could be extracted from the measured methanol concentrations **(Table S5)**. The permeability at 1M methanol concentration was similar for all graphene samples and in the range of 12 - 14e$^{-7}$ cm$^2$ s$^{-1}$, less than half of that of the Nafion® reference at 32.4e$^{-7}$ cm$^2$ s$^{-1}$. At 5M methanol, the permeability depended on the graphene loading. The minimum of 12.5e$^{-7}$ cm$^2$ s$^{-1}$ was achieved for the 130 µg cm$^{-2}$ graphene sample, which is about 20 times lower than the permeation through the Nafion® reference of 243.5e$^{-7}$ cm$^2$ s$^{-1}$ at 5M methanol concentration.

The proton conduction mechanism in PFSA membranes is well explored [18]. There is wide agreement that proton transport in well hydrated membranes is dominated by the Grotthus-mechanism [18,19], which states that water molecules adsorb at the sulfonic-acid sides of the PFSA polymeric chains and present a vehicle for proton-hopping. During this process, protons travel between adjacent water molecules and thus migrate from anode to cathode. The proton conduction mechanism in layered carbonaceous materials like graphene oxide is also assumed to be governed by a Grotthus-mechanism [20]. According to Karim et al., the hydrophilic oxygen-rich functional groups of the graphene oxide leads to the formation of a water film between graphitic planes, which then facilitates the proton movement along the water film [20]. The electrochemically exfoliated graphene used in this work has hydrophilic functional groups at the flake edges. Although these have not been investigated in detail, it is likely that water molecules can adsorb at those sites and play a similar role as oxygen-rich groups in graphene oxide.

The reduced fuel permeability found in the hybrid membranes can be explained by a blocking mechanism, which takes place in the graphene flake network. Protons can move along the graphitic planes, i.e. parallel to the substrate, and perpendicular through gaps in the flake network by the Grotthus-mechanism. Larger ions and molecules, however, are likely prohibited from crossing the layers in this way. The interlayer spacing of the percolated graphene flake network plays an important role concerning its barrier properties. Other geometrical factors like flake size and distance of adjacent flakes can affect the barrier as well. In 2016, Cheng et al. demonstrated free standing layered graphene gel (LLG) membranes with variable interlayer spacing [21]. They proposed a transport mechanism of ions ($K^+$, $Cl^-$) permeating through these graphene-based nano-networks. In their model, a labyrinth like porous structure is considered, whose pathways present an increase of travel distance that each ion needs to pass in order to permeate through the membrane. A set of geometrical variables is used to describe the layer structure. The lateral flake size $L$, the distance between neighboring flakes $\delta$ and the average interlayer spacing $d$ are identified as key parameters for ionic transport. This model is also applicable for the $H^+$ conductivity through the coating layers presented in this work and supported by the cross-sectional STEM images in **Figure 2c**, as those confirmed a barrier network of parallelly aligned graphene flakes in line with the model.

In addition to the variation of layer thicknesses, the influence of different flake sizes on the barrier properties was studied, a matter that is controversially discussed in literature. In the previous section, a graphene loading of 130 μg $cm^{-2}$ showed the best performance, so this amount was kept constant. I-V-plots of the corresponding MEAs at 1M and 5M methanol feed concentrations are shown in (**Figure 3b**). The voltage breakdown of the MEA with 300 nm flakes (red data points) happens at lower current densities than for the MEA with 1 µm flakes (blue) and the pure Nafion® N115 reference (black). This is true for both 1M and 5M methanol feed concentrations. Our data from flakes with mean lateral sizes of 300 nm and 1 µm suggests the use of larger flakes in order to maximize the barrier effect. Those findings support the hypothesis that larger flakes would lead to longer transversal/in-plane pathways for permeates and thus reducing permeability [10,22]. Note that contradictory results are published as well [21,23].

We observed a clear influence of the methanol feed concentration on the fuel cell performance: the graphene coated samples exhibit higher DMFC power densities at elevated methanol feed concentrations (e.g. 5M) compared to a Nafion® N115 reference, whereas their performance is lower at 1M methanol feed concentration. The findings suggest that for low methanol concentrations the proton conductivity dominates the achievable DMFC power density, while for higher concentrations the methanol permeability is critical. This is in line with previous data published by Lin and Lu in 2013 [11]. They observed that GO-laminate on Nafion® exhibits superior DMFC performance compared to a Nafion® reference at 6M and 8M methanol feed concentration, while being inferior for lower concentrations. In this discussion, it is important to observe that Nafion® membranes are known to swell with increasing methanol concentrations [4]. Thus, the conductive channels in Nafion® widen and promote the permeation of methanol molecules. Accordingly, we measured a nearly ten-fold increase of methanol permeability on our Nafion® N115 reference samples at 5M ($243.5e^{-7}$ cm² s$^{-1}$) compared to the values at 1M ($32.4e^{-7}$ cm² s$^{-1}$), in line with previously reported measurements [10]. In contrast, graphene coated Nafion® with 130 µg cm$^{-2}$ loading exhibits a nearly constant permeability of 12.3e$^{-7}$ cm² s$^{-1}$ at 1M and 12.5e$^{-7}$ cm² s$^{-1}$ at 5M. Our data is summarized and benchmarked in **Table 1**. We consider the graphene interlayer spacing in a different way than the conductive channels in Nafion®. We base this consideration on You *et al.* [24], who *in-situ* measured the interlayer spacing of graphene oxide laminate soaked in a water/methanol binary solutions. They found the highest interlayer spacing for pure water (10.4 Å), and observed a decrease of the spacing for increasing methanol parts (<9 Å, 20-100 % v/v methanol). We therefore conclude that the interlayer spacing of graphene flakes depends on the methanol concentration and that it has a major influence on the permeability of the Nafion®/graphene-samples.

## Conclusions

Homogeneous, percolated graphene thin films of highly aligned, electrochemically-exfoliated graphene (e-G) flakes were deposited on Nafion® membranes by a simple and scalable spray coating process to enhance their selective diffusion barrier properties against fuel crossover. The graphene coatings efficiently reduced the hydrogen permeability of Nafion® HP membranes up to factors of 39 in a dry state at 80°C and 29 at 80°C and 65% relative humidity. Furthermore, the methanol permeability at 5M feed concentration was reduced almost 20 times compared to a Nafion® N115 reference sample. We found that the lateral graphene flake size plays an important role for the fuel blocking effect, with the best DMFC performance observed for the largest flakes of about 1 µm in size. The coated membranes facilitate DMFC operation at high methanol concentrations as they ensure a significantly decreased methanol permeability without significantly reducing the intended proton conductivity. At 5M methanol, the maximum power density is improved to 390% of that of bare Nafion® N115 reference samples, a surpassing value to the best of our knowledge. In summary, the developed graphene coatings are an efficient measure for reducing the fuel crossover and for enhancing the performance of fuel cells, both DMFCs as well as PEMFCs, that may be easily integrated in membrane electrode manufacturing lines.

**Experimental methods**

Nafion® membranes were covered with dense, continuous graphene thin films by spray coating of customized graphene dispersions based on G-DISP-H2O-CSO by Sixonia Tech GmbH. This commercially available dispersion is a water-based formulation without surfactants or other additives. It contains electrochemically exfoliated, functionalized few-layer graphene flakes. Utilizing different sonication and subsequent centrifugation procedures, dispersions with different lateral flake sizes were prepared with the mean sizes ranging from approximately 200 nm to 1 µm. The general fabrication process of the product is described in ref. [25] and a review of the electrochemical exfoliation process can be found in ref. [26]. The flake concentration was adjusted to 1 g $L^{-1}$ for further processing by dilution with DI-water. Commercially available Nafion® HP and Nafion® N115 membranes were used as substrates and references. The samples (70 x 70 $mm^2$) were coated by spray deposition using a Badger 200 NH airbrush gun and compressed air as the propellant gas. The amount of deposited graphene was varied from 25 to 150 µg $cm^{-2}$. After the deposition process, the samples were dried at ambient conditions.

A ZEISS Supra 60VP FE-SEM was used to analyze the surface structure of the graphene layers (in-lens detector). Raman spectra of the deposited films were recorded using a WITEC 300R system at a laser wavelength of λ = 532 nm and a laser power of 3 mW. The sample surface morphology was characterized by AFM (Veeco Instruments, Bruker) before and after graphene deposition. Cross-sectional preparation of electron transparent TEM samples was conducted by ultramicrotomy using a Leica EM UC7 microtome. To conserve the morphology of the membrane-graphene stack as well as the complete MEAs/fuel cells, small pieces of the samples were extracted by doctor-blade cutting and were embedded in epoxy (Epofix, Struers, Germany). Curing of the epoxy was achieved by annealing at 60 °C for 4 h. A truncated pyramid with a trapezoid-shaped block face was trimmed (DiATOME trim knife) to reach the relevant sample regions. Cross-sectional TEM samples with a thickness of approximately 50 nm were cut with a DiATOME diamond knife and subsequently transferred onto Lacey carbon TEM grids (TedPella/Plano) for enhanced support. Transmission electron microscopy was performed utilizing a Thermo Fisher FEI Talos F200X operated at 200 kV. The instrument is equipped with a XFEG high-brightness gun and a Gatan Continuum ER spectrometer (with high-speed DualEELS, dispersion 0.05 eV/channel, DigiScan, EDXS integration, GMS 3.4x) for electron energy-loss spectroscopy (EELS). ADF-STEM imaging was conducted for morphological and structural analyses, while advanced EELS was employed for detailed chemical-bond analyses. Dual EELS was used for simultaneous recording of core-loss and corresponding low-loss spectra to systematically recalibrate the energy-loss and to remove multiple scattering from the core-loss spectra by deconvolution (relative sample thickness around 0.4–1.2 mean-free paths for inelastic scattering depending on the local material). Detailed chemical-bond analyses were carried out for three major components, namely the epoxy, Nafion® and the graphene layers. Reference spectra were obtained from homogeneous areas of the mapping itself by summing up respective spectra. Each individual spectrum of a mapping was evaluated by multiple least-squares of those references to the carbon ionization edge in the energy-loss range of 280–315 eV, which served to distinguish between the different components.

The membrane performance was evaluated in fuel cell operation by assembling MEAs with graphene-G coated Nafion® membranes and bare Nafion® reference samples. The catalyst layers were applied by an ultrasonic nozzle (Exacta Coat FC Fuel Cell Coating System, Sono-Tek, USA) while the substrate was fixed on a heated vacuum-plate. The cathode catalyst layer was deposited on the bare side of the membrane, while the anode catalyst layer was sprayed on the graphene coated side. A detailed description of the MEA assembly process can be found in ref.[27]. DMFC MEAs were tested in a fuel cell setup qCf FC25/100 V1.1 (Baltic FuelCells GmbH) with an electrode geometrical area of 25 $cm^2$ and liquid cooling/heating. DMFC polarization curves were acquired by applying specific current loads to the measured cell (galvanostatic method). Through-plane proton conductivities were measured in the same setup utilizing electrochemical impedance spectroscopy (EIS). Methanol permeation experiments were performed likewise in the same setup, but under ex-situ operating conditions. At the anode side, a constant methanol-water premix was supplied as usual but at the cathode side a constant water flow instead of an air feed was used. The concentration of the permeating methanol in the cathode outlet was monitored by a liquid methanol sensor (microMCS, ISSYS), the methanol diffusion flux was measured in equilibrium state. The hydrogen permeation was measured ex-situ using another setup qCf FC25/100 V1.1 connected to a fuel cell test rig in $H_2/N_2$ modus to realize correct humidity conditions. A mass spectrometer (MS) OmniStarTM GSD 320 O1 (Pfeiffer Vacuum GmbH) was employed to determine the concentration of permeating hydrogen.

**Associated content**

**Supporting Information**

Measurement setups, SEM, STEM, EELS, Raman, methanol permeabilities.

**Author information**

**Notes**

**Acknowledgements**

The authors thankfully acknowledge funding through German Ministry for Economic Affairs and Energy BMWi (PROTONLY, 49VF170038). Part of this work was performed at the DFG-funded Micro-and Nanoanalytics Facility (MNaF) of the University of Siegen (INST 221/131-1) utilizing its major TEM instrumentation (DFG INST 221/93-1, DFG INST 221/126-1) and sample preparation equipment. JR, UP, MM and MCL acknowledge discussions within the Aachen Graphene and 2D Materials Center.

## Figures

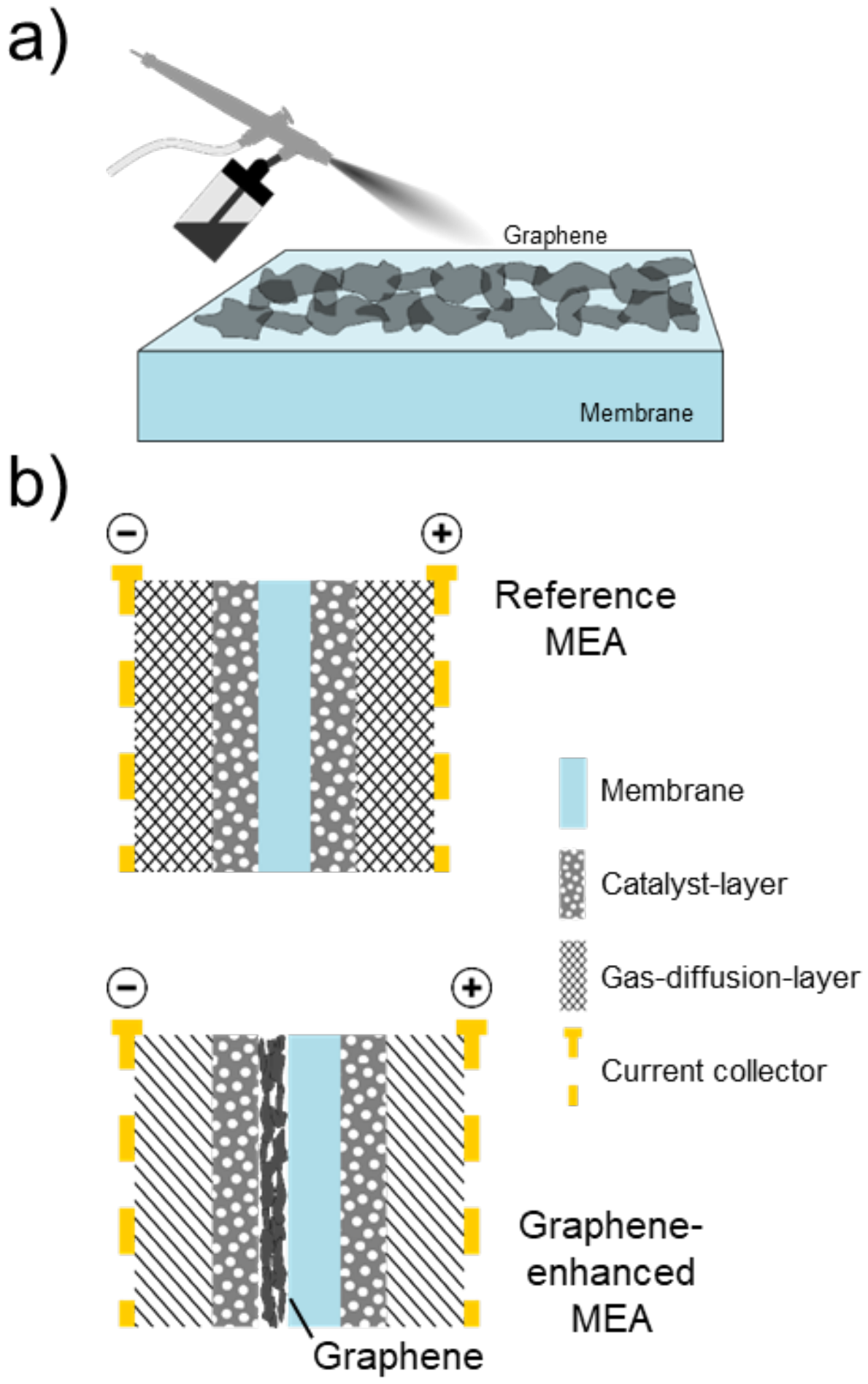

**Figure 1.** (a) Schematic illustration of the sample fabrication by spray-coating. (b) Cross-sectional schematic of two MEAs comparing a reference MEA (top) and a graphene-enhanced MEA (bottom).

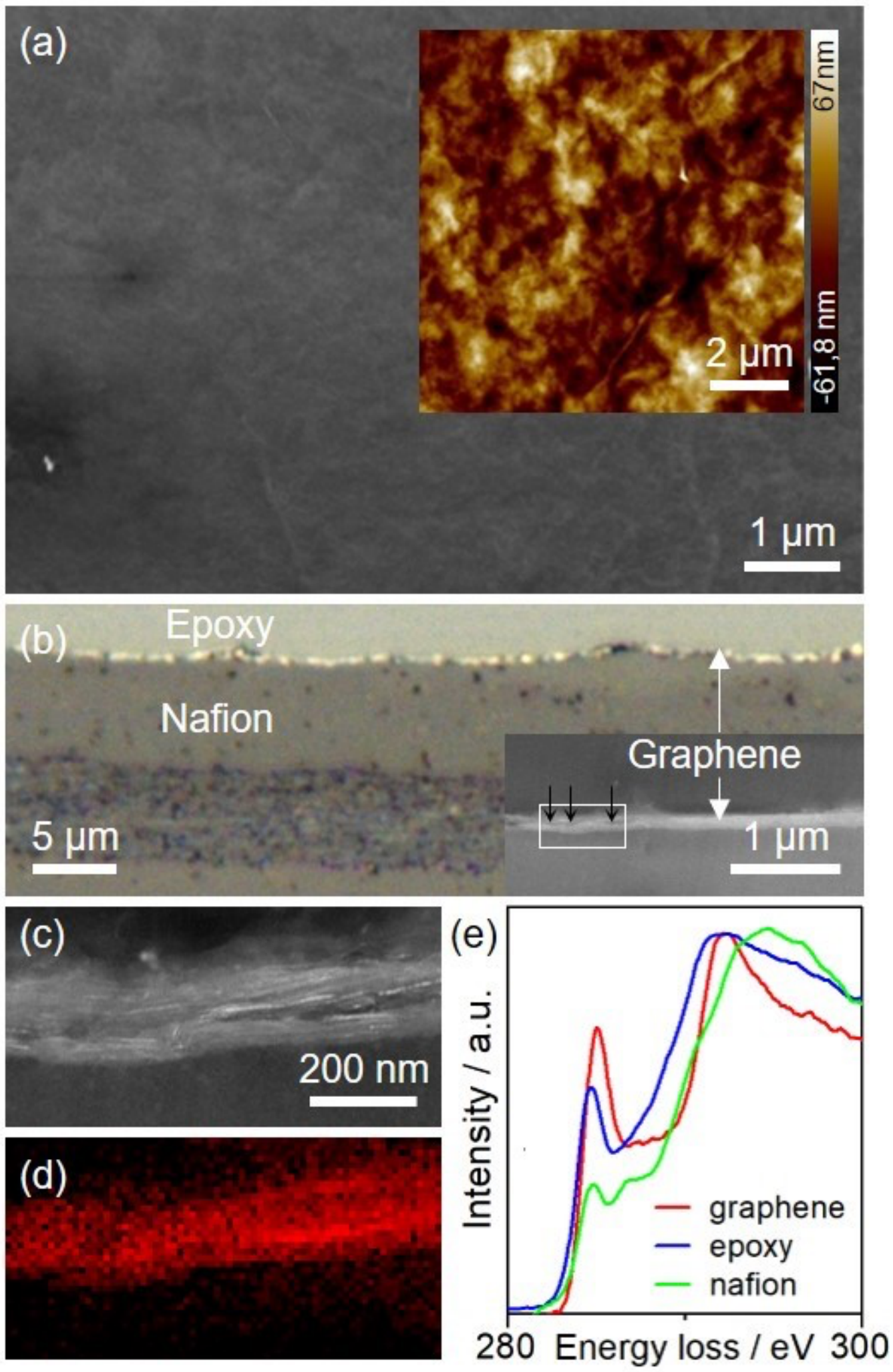

**Figure 2.** (a) SEM top-view image observed under 29° tilting angle. The layer is smooth and densely packed, and single flakes are difficult to distinguish. Inset: topography plot measured by AFM. (b) Optical micrograph of the cross-section of the graphene-Nafion® layer system embedded in an epoxy matrix. Inset: Scanning transmission electron micrograph acquired in annular dark-field mode (ADF-STEM) of the cross-section. The graphene layer features the highest intensity in the image. Black arrows indicate flaws in the otherwise dense/smooth graphene film. (c) ADF-STEM image of the labeled area in inset of (b) Individual graphene flakes are visible as bright lines. (d) Respective multiple linear least-squares (MLLS) fit of the internal graphene reference (red line in (e)) to the recorded electron energy-loss mapping of the same area as in (c). (e) EELS reference spectra of the individual components in the cross-sectional sample derived from the locally integrated STEM-EELS spectra from the respective mapping regions (after background subtraction).

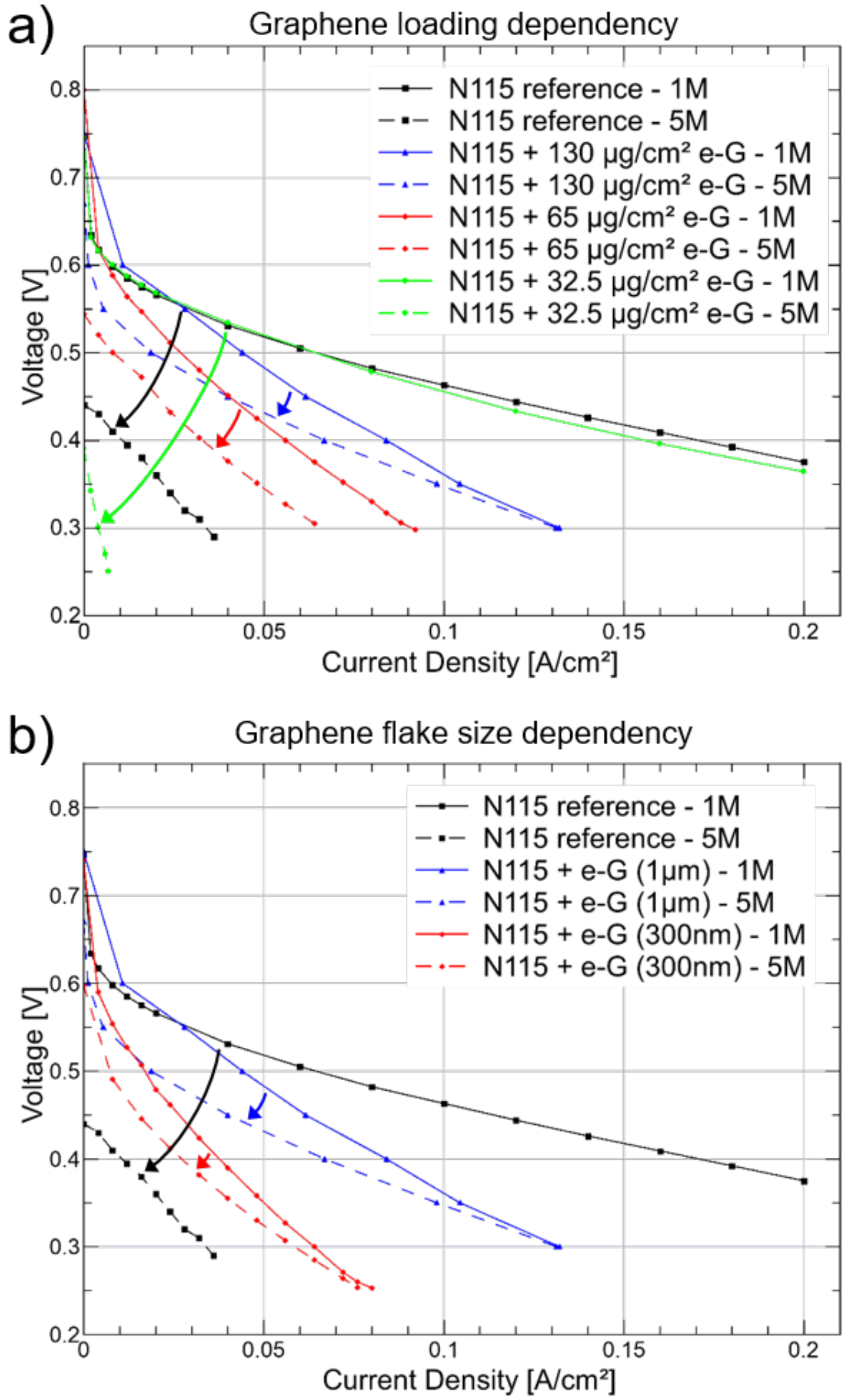

**Figure 3.** Effect of graphene loading and graphene flake size on DMFC performance: The cell output voltage was plotted versus a variable current density, which was applied to the fuel cell during operation. I-V-polarization curves of Nafion® N115 reference membrane were compared to graphene coated N115 (a) with various amounts of graphene and (b) with different graphene flake sizes at a constant graphene loading of 130 µg cm$^{-2}$. The arrows indicate the change of fuel cell performance by increasing the methanol concentration from 1M to 5M.The MEA with 130 µg cm$^{-2}$ graphene and 1 µm flake size shows superior performance for 5M methanol compared to the reference. 1 µm flakes show better performance compared to the 300 nm sample. This indicates that larger flakes (lateral size) perform better as barrier layer

## Tables

**Table 1.** Comparison chart of various Nafion-based DMFC membranes.

| Additive | Membrane configuration and fabrication | Proton conductivity [S cm$^{-1}$] | Methanol permeability [cm$^2$ s$^{-1}$] x $10^{-7}$ | Max. Power Density [mW cm$^{-2}$] | Nafion® reference [mW cm$^{-2}$] | Relative improvement [%] | Ref. |
|---|---|---|---|---|---|---|---|
| - | Commercial Nafion® N115 (reference) | 0.093 [a)] | 243.5 @ 5M [c)] | 10 @ 5M, 65°C | - | - | This work |
| e-G | Nafion® N115 with e-G coating | 0.078 [a)] | 12.5 @ 5M [c)] | 39 @ 5M, 65°C | 10 @ 5M, 65°C | 390 | This work |
| zeolite | Nafion®/zeolite-composite | 0.088 [b)] | 14 @ 2M [d)] | 120 @5M, 70°C | 62 @5M, 70°C | 193 | 28 |
| zeolite | Nafion® with Nafion®/zeolite-composite coating | 0.09 [a)] | 45 [e)] | 48 @4M, 70°C | ~22 @4M, 70°C | ~224 | 29 |
| GO | Nafion®/GO/Nafion®-stack | 0.004 [b)] | 0.32 @ 5M [d)] | 51 @5M, 60°C | ~33 @5M, 60°C | ~155 | 10 |
| GO | Nafion®/GO-paper-laminate | 0.024 [a)] | 9.3 @ 0.75M [d)] | 55 @6M, 50°C | 40 @6M, 50°C | 138 | 11 |

[a)] Electronic impedance spectroscopy (EIS), through-plane
[b)] Four electrode impedance measurement, in-plane
[c)] Working cell direct measurement, constant flow
[d)] Two compartment diffusion cell method
[e)] Working cell indirect measurement, constant flow

**For Table of Contents Only**

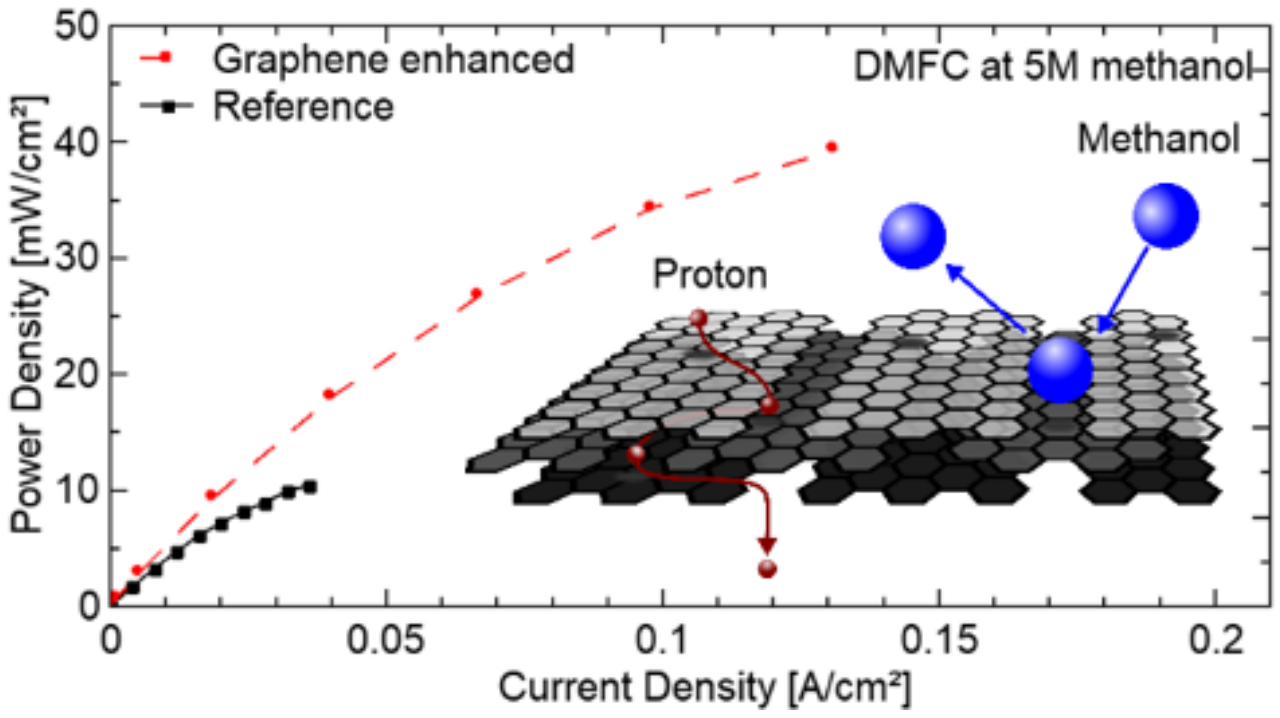

Supporting Information

## Graphene coating of Nafion® membranes for enhanced fuel cell performance

*Jasper Ruhkopf, Ulrich Plachetka, Michael Moeller, Oliver Pasdag, Ivan Radev, Volker Peinecke, Marco Hepp, Christian Wiktor, Martin R. Lohe, Xinliang Feng, Benjamin Butz, Max C. Lemme**

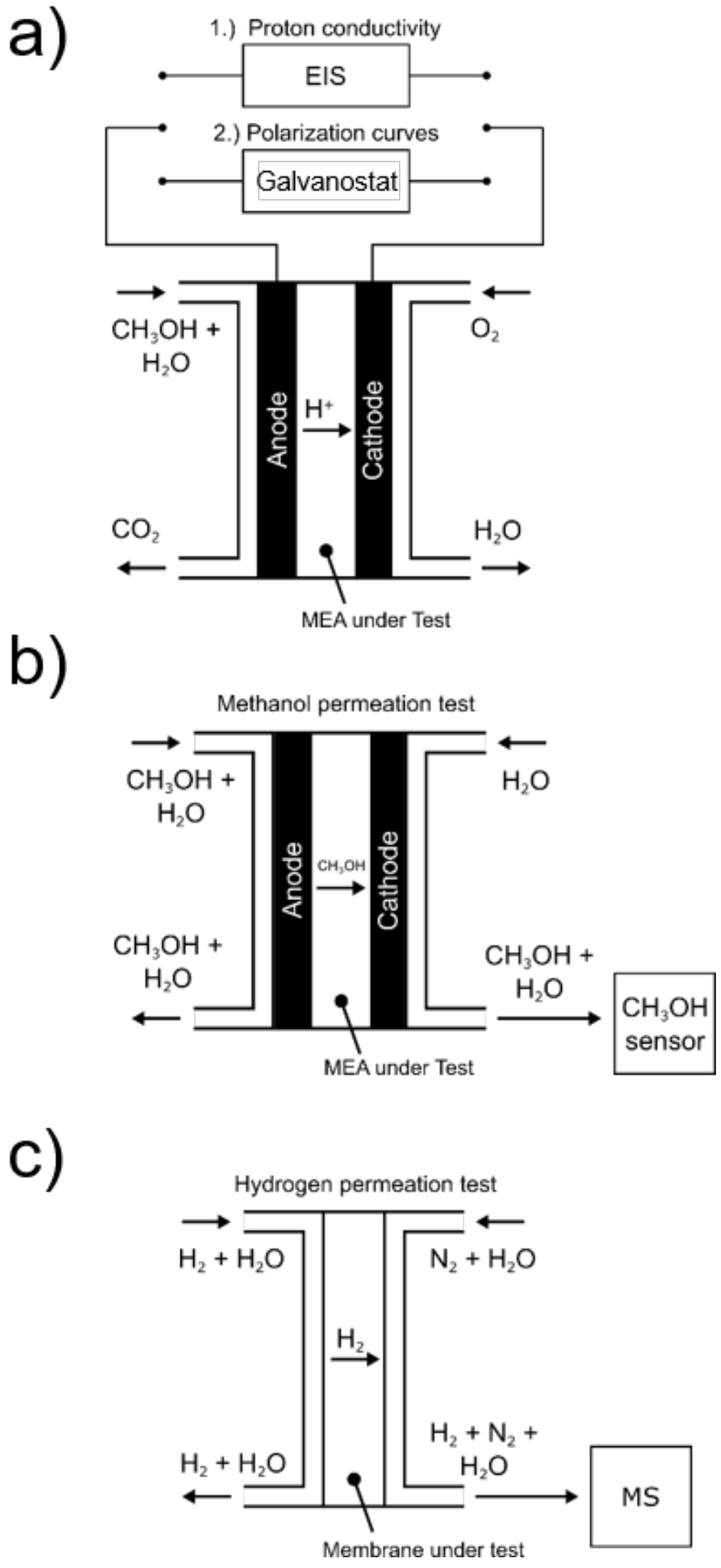

**Figure S1**. Measurement setup schematics. (a) Measurement of proton conductivity in fuel cell test rig with electrochemical impedance spectroscopy (EIS) and recording of polarization curves by galvanostatic method (b) Measurement of methanol permeation with liquid methanol sensor. (c) Hydrogen permeation measurement through membrane in dedicated cell. A mass spectrometer (MS) is used to analyze hydrogen concentration at cathode outlet.

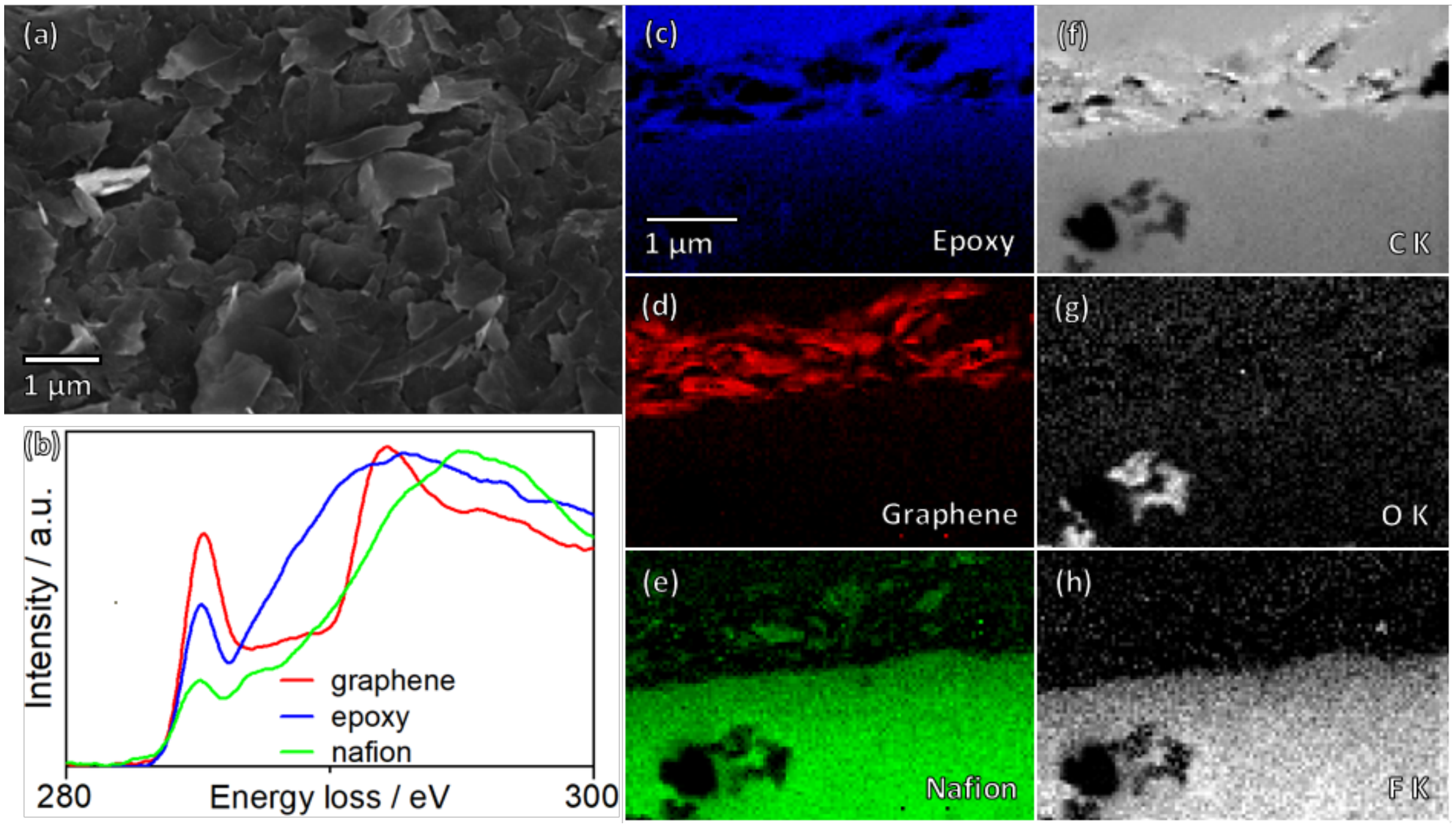

**Figure S2.** Supporting data of a previous coating attempt using commercial graphene nano-platelet material (a) SEM top-view image observed under 29 ° tilting angle. Individual flakes can be distinguished and do not form a smooth and closed film. (b) Smoothed EELS reference spectra for graphene, the epoxy resin and Nafion created from the locally integrated STEM-EELS signal after background subtraction. (c-e) Respective MLLS fits of the internal references shown in (b) to the recorded STEM-EELS spectrum image. (f-h) Elemental distributions of C, O and F based on their respective K edges in the according STEM-EELS spectrum image.

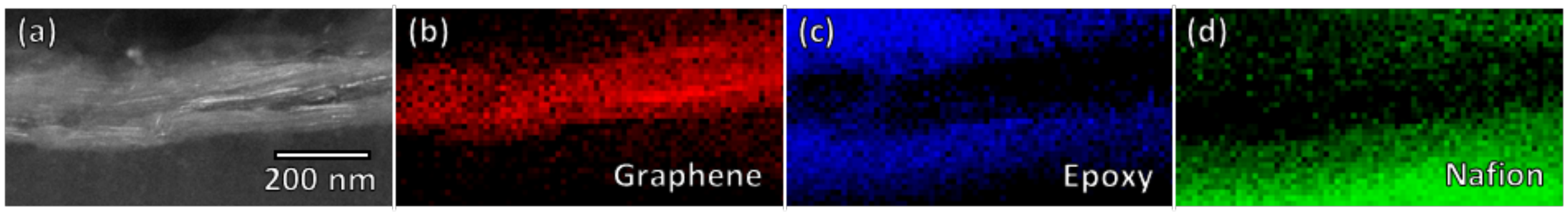

**Figure S3.** (a) ADF-STEM cross-sectional micrograph (same as Fig. 2c). (b-d) Respective multiple linear least-squares (MLLS) fits of the spectral graphene, Nafion, and epoxy references (shown in Fig. 2e) to the EELS mapping of the same area as in a).

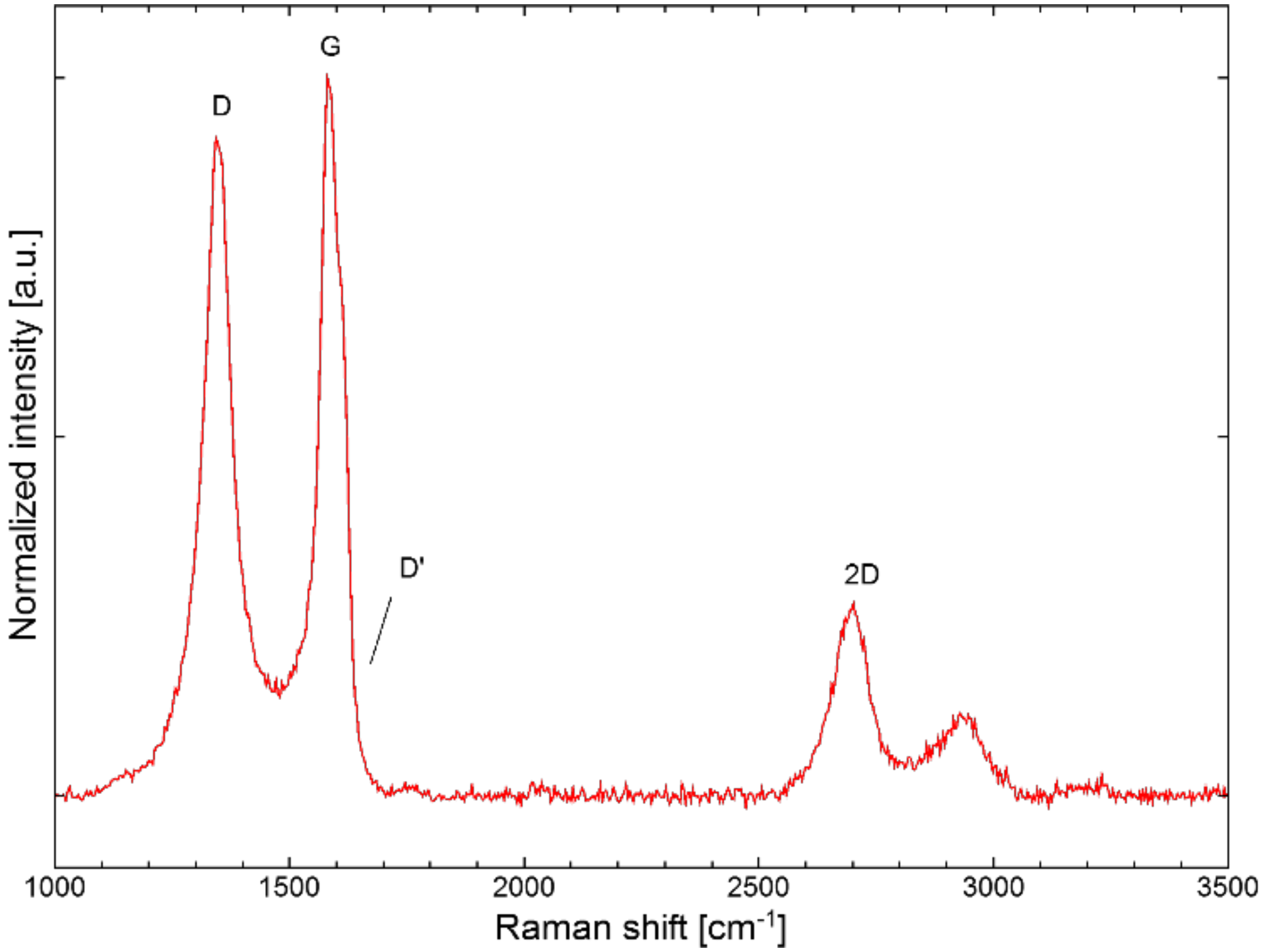

**Figure S4.** Raman spectrum of e-G thin film, normalized to G-peak.

**Table S5.** Methanol Permeabilities of graphene-coated membranes, measured at 1M and 5M MeOH concentration.

| Membrane | MeOH Permeability [$cm^2$ $s^{-1}$] x $10^{-7}$ | |
|---|---|---|
| | 1M MeOH | 5M MeOH |
| Nafion® N115 reference | 32.4 | 243.5 |
| N115 + 32.5 µg $cm^{-2}$ e-G | 14.3 | 129.4 |
| N115 + 65 µg $cm^{-2}$ e-G | 12.9 | 108.2 |
| N115 + 130 µg $cm^{-2}$ e-G | 12.3 | 12.5 |